%% file: main.tex
\documentclass[sigconf]{acmart}

\PassOptionsToPackage{bookmarks=false}{hyperref}
\hypersetup{bookmarksdepth=-2}

\usepackage[ampersand]{easylist}

\setcopyright{rightsretained}

\acmConference[ICSE 2018]{ICSE}{June 2018}{Gothenburg, Sweden} 
\acmYear{2018}
\copyrightyear{2018}

\acmArticle{4}
\acmPrice{0.00}

\begin{document}

\copyrightyear{2018} 
\acmYear{2018} 
\setcopyright{acmcopyright}
\acmConference[SE4COG'18]{SE4COG'18:IEEE/ACM 1st International Workshop on Software Engineering for Cognitive Services }{May 28--29, 2018}{Gothenburg, Sweden}
\acmBooktitle{SE4COG'18: SE4COG'18:IEEE/ACM 1st International Workshop on Software Engineering for Cognitive Services , May 28--29, 2018, Gothenburg, Sweden}
\acmPrice{15.00}
\acmDOI{10.1145/3195555.3195567}
\acmISBN{978-1-4503-5740-1/18/05}

\title{Smart Conversational Agents for Reminiscence}


\author{Svetlana Nikitina}
\affiliation{%
  \institution{University of Trento, Italy}
}
\affiliation{%
\institution{Tomsk Polytechnic University}
}
\email{svetlana.nikitina@unitn.it}

\author{Sara Callaioli}
\affiliation{%
  \institution{University of Trento, Italy} 
}
\email{sara.callaioli@studenti.unitn.it}

\author{Marcos Baez}
\affiliation{%
  \institution{University of Trento, Italy}
}
\affiliation{%
\institution{Tomsk Polytechnic University}
}
\email{baez@disi.unitn.it}




\begin{abstract}
In this paper we describe the requirements and early system design for a smart conversational agent that can assist older adults in the reminiscence process. The practice of reminiscence has well documented benefits for the mental, social and emotional well-being of older adults. However, the technology support,  valuable in many different ways, is still limited in terms of need of co-located human presence, data collection capabilities, and ability to support sustained engagement, thus missing key opportunities to  improve care practices, facilitate social interactions, and bring the reminiscence practice closer to those with less opportunities to engage in co-located sessions with a (trained) companion. We discuss conversational agents and cognitive services as the platform for building the next generation of reminiscence applications, and introduce the concept application of a smart reminiscence agent.


\end{abstract}

%
%



\maketitle

\input{sections/introduction}

\input{sections/related}

\input{sections/requirements}
\input{sections/design}

\input{sections/architecture}
\input{sections/conclusion}

\begin{acks}
This work has received funding from the EU Horizon 2020 research and innovation programme under the Marie Sk\l{}odowska-Curie grant agreement. It was also supported by the project "Evaluation and enhancement of social, economic and emotional wellbeing of older adults" under the agreement No.14.Z50.31.0029, Tomsk Polytechnic University.
\end{acks}

\bibliographystyle{ACM-Reference-Format}
\bibliography{reminiscence,chatbots} 

\end{document}

%% file: sections/introduction.tex
\section{Background and Motivation}
In this paper we define the requirements and early design of a conversational agent that supports engaging and effective conversation about a person's life and memories. 
Specifically, we discuss the case of personal and social \textit{reminiscence} sessions by older adults.

Reminiscence is the process of collecting and recalling past memories through pictures, stories and other \textit{mementos} \cite{webster2007reminiscence}. 
The practice of reminiscence, and more generally, reminiscence therapy and life review, have well documented benefits on 
emotional and mental wellbeing \cite{bender1998therapeutic,subramaniam2012impact}, 
interpersonal relationships and interactions \cite{huldtgren2015probing,allen2009legacy}, 
and preserving personal identity \cite{crete2012reconstructing,fivush2011making}. 
These benefits, along with the potential of reminiscence to transfer knowledge, stimulate conversations, reduce boredom, and maintain intimacy \cite{webster1993construction} makes it a very desirable practice, especially for older adults.


Research on technology-mediated reminiscence have taught us valuable lessons on how to facilitate usage by older adults \cite{west2007memento,piper2013audio}, collect memories and support storytelling \cite{lee2016picmemory,lee2014picgo}, stimulate cognitive functions \cite{subramaniam2016digital,lee2014picgo}, and support conversations \cite{kim2006cherish,astell2009working}. However, these tools are very limited when it comes to creating opportunities for social interactions and for supporting social integration, something that ample and concordant research tells us it is a determinant of wellbeing at any age \cite{hartup1999friendships}.
In our previous work, we addressed some of these limitations in a reminiscence-based social interaction tool \cite{ibarra2017stimulating} that aimed at stimulating conversations and creating bonds among older adults in residential care. However, our previous efforts and those of the community have reached a limit in terms of what traditional reminiscence technology can do to support older adults. 

As we will see, traditional reminiscence technology 
i) strongly relies on co-located human presence for assisting in the reminiscence process, engaging in conversations or joining social reminiscence sessions; this often requires dedicated personnel, the presence of family member and friends, which greatly limit the practice and potential benefits of reminiscence, especially for those with less access to social contacts; 
ii) do very little to actively guide reminiscence sessions in a way that is effective and engaging, requiring users (the participant or a guide) to recall important aspects of the participant's life, to rely only on intentional triggers, or to follow predefined templates; thus missing the opportunity to reflect interests and previous stories to make sessions more fun and satisfying, and   
iii) memory collection and digital storytelling is mostly limited to archiving and browsing, missing the potential benefits of building a rich profile of users, e.g., to improve care practices \cite{thorgrimsdottir2016reminiscence} or facilitate social interactions \cite{sanchiz2016makes,caforio2017viability}, for example by identifying people with common values, or simply by making one's own family, and especially grandchildren, aware of the rich and often interesting life their grandparents have lived and are living.



Conversational agents and the ever growing number of cognitive services offer a unique platform to engineer the next generation of smart reminiscence systems that are not only more personal and engaging, but that can make those who have less opportunities for co-located interactions enjoy the benefits of the reminiscence practice. Engineering such a system requires insights into the reminiscence process, the development conversational models for reminiscence, the underlying cognitive services and system design.

In this short paper we describe the requirements, models and a system design for such a conversational agent. From a technological standpoint we investigate the design of a chatbot that can flexibly drive a person (or a family) through a picture-driven storytelling session. The goal of the bot is to be able to sustain a conversation about a person's life, while harvesting useful information and while keeping the session interesting and engaging.

%% file: sections/related.tex
\section{Related work}


Exploring the use of technology to support the reminiscence practice by older adults has been subject of plenty of research (see \cite{lazar2014systematic} for a review on the subject). The work on this area can be summarised in efforts to facilitate usage by older adults \cite{west2007memento,piper2013audio}, collect memories and support storytelling \cite{lee2016picmemory,lee2014picgo}, stimulate cognitive functions \cite{subramaniam2016digital,lee2014picgo}, and support conversations \cite{kim2006cherish,astell2009working,ibarra2017stimulating}. 
Despite the valuable contributions, most of these works still rely either on the ability of participants to drive the personal reminiscence sessions or the presence of assistance \cite{astell2009working,ibarra2017stimulating}. 
Support for social interactions is also mostly focused on co-located settings \cite{uriu2009caraclock,astell2009working} or offered in the form of online sharing \cite{west2007memento}. 
The efforts in improving data collection are also largely focused on collaboration \cite{lee2016picmemory} and optimising the interaction design \cite{lee2014picgo}, with little to no intelligence. This represents not only a missed opportunity but also a constraint for those people with less opportunities for face to face interactions. 

Another relevant type of technology is that of digital companions and relational agents. 
The feasibility of social companions for older adults, exploring a wide ranging set of activities was evaluated in \cite{vardoulakis2012designing}, using the Wizard of Oz technique, with a researcher selecting the agent responses for a period of a week. A interesting finding from this  work is identifying "storytelling" as the type of interaction elders spent more time with the agent. 
In a follow up study \cite{ring2013addressing}, the authors explored the effects of proactive vs passive behavior in conversational agents using a scripted conversation system, identifying that agents capable of initiating interactions  (proactive) were more effective in addressing loneliness. 

Similarly, a commercial digital companion driven by a human operator was also evaluated in \cite{demiris2016evaluation}. The companion, an always on digital pet with conversational skills, was positively assessed by the participants, especially in the ability to show pictures (sent by a caregiver) and "remember" things about the older adults (e.g., a scripted reminder that included the birthday of the participant).  

Another thread of work focused on identifying requirements as well as older adults attitudes toward autonomous virtual agents (not driven by a human operator). Tsiourti et al. \cite{tsiourti2014virtual} explored the attitudes and perception of older adults towards different aspects of virtual companions during the design phase, pointing to the "usefulness" and "social intelligence" a determinant factors for the success. The authors later conducted a exploratory study in 20 homes \cite{tsiourti2016virtual} with a virtual companion capable of assisting every day tasks, identifying several challenges, especially in the mismatch between expectations of the older adult in what regards the conversational capabilities of the virtual companion. 


What the above tells us is that the use of conversational agents is feasible, and that reminiscence  is an untapped and desirable application to explore. We also see that the success in previous studies was evident when agents displayed qualities attributed to humans and when proved useful to the task at hand -- not surprisingly this was the case in agents operated by humans. 
All things considered, the use of cognitive services in reminiscence technology still remains in its infancy.  
This clearly motivates the need for research and development of intelligent conversational agents can assist the reminiscence process.




%% file: sections/requirements.tex
\section{Requirements}

To summarize the extensive theoretical work on the subject into a practical reference framework for technology-mediated reminiscence, we take the conceptual reminiscence model by Webster et al. \cite{webster2010mapping} and instantiate it in a process that captures the most salient technology roles from the literature. 

\begin{figure}
\centering
\includegraphics[width=\columnwidth ]{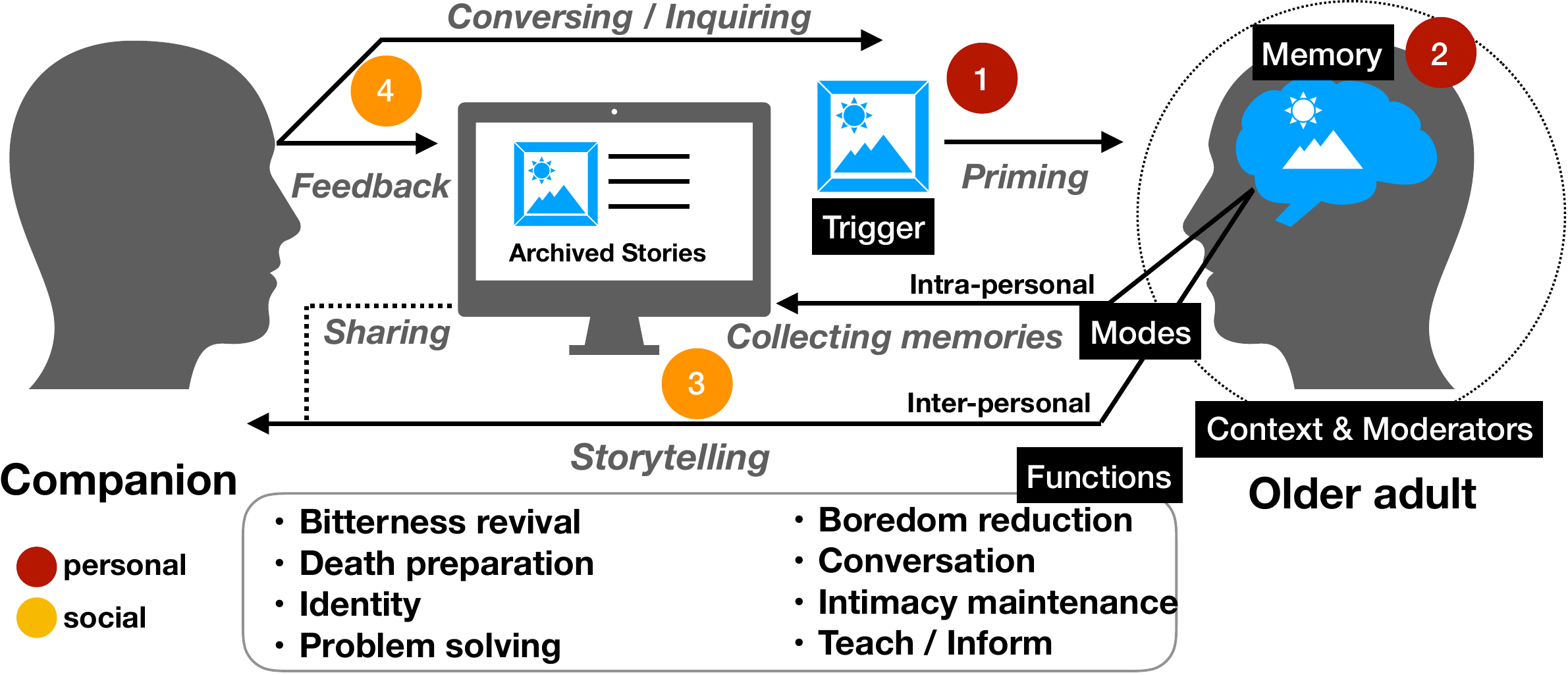}
\caption{Technology-mediated reminiscence process}
\label{fig:rem-process}
\end{figure}

This summary, illustrated in Figure \ref{fig:rem-process}, shows how the process is initiated by a \emph{trigger}, usually a memento in the form of pictures, videos or music, or specific questions. The trigger results in memories being primed, which can be then shared (virtually or face to face) via storytelling, or used for personal reflection. These two visibilities of memories define two distinctive \emph{modes} known as interpersonal and intrapersonal. Memories and related stories resulting from this elicitation process can be digitally collected. If memories are used for interpersonal reminiscence in a storytelling session, then companions can engage in conversations, provide feedback, and further inquire about the aspects of the stories told by the person reminiscing. The role of the companion can vary depending on the degree of participation: companions can take a leading role in guiding the reminiscence session, co-participate in the case of social reminiscence, or take a more passive role as recipient of the storytelling. 
Regulating the entire process we have the \textit{context and moderators}. These are intrinsic properties of the person (and reminiscence session) that can influence the reminiscence process and outcome, such as aspects of personality, age, gender, and relationship with the reminiscing companion. Finally, the practice of reminiscence is related to very specific functions, among which we highlight: i) boredom reduction, referring to the potential to create joyful experiences, ii) conversation, to the opportunity to create a dialog, iii) and teach / inform, to the opportunity to transfer stories and knowledge. 


\begin{figure*}
\centering
\includegraphics[width=\textwidth ]{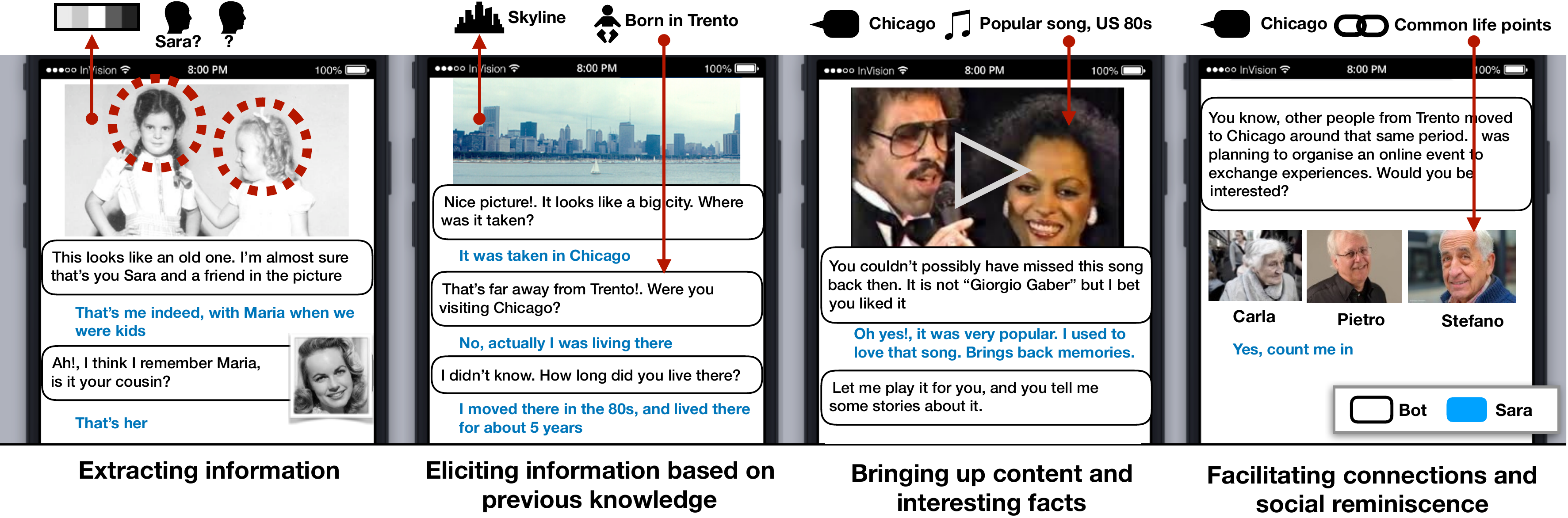}
\caption{Example conversations illustrating chatbot skills}
\label{fig:bot-mockup}
\end{figure*}

Informed by the above process, and our own experience designing and evaluating reminiscence technology for nursing home residents (on average, 80+ older adults that require assistance to perform activities of daily living) \cite{ibarra2017stimulating}, we identified the following main requirements:

\begin{easylist}
\ListProperties(Hide=100, Hang=true,Style*=\textbullet~)
&  \textbf{Eliciting memories},  in a way that engage participants in storytelling, in conversations that are sensitive to (changing) interests of the user, the topic, sentiment, and context. 


&  \textbf{Archiving life stories and memories}, to support personal reflection, knowledge transfer or even virtual storytelling. 

&  \textbf{Building representations of the life of a person}, not only to facilitate browsing, but to serve the growing need for historical information about older adults, e.g., to improve care practices \cite{thorgrimsdottir2016reminiscence}, facilitate social interactions \cite{sanchiz2016makes,caforio2017viability}, and build better context and moderators to improve the reminiscence process.

&  \textbf{Navigating interests and passions}, as to engage users in conversations where they can have fun by navigating their passions (e.g., music, video, movies).  This also helps elicit stories, and more importantly collecting preferences.

&  \textbf{Facilitating (re-)connections}, by identifying potential companions and creating opportunities for social reminiscence and interactions. Interactions are facilitated to avoid stress in initiating conversations. 


\end{easylist}

In addressing the above requirement, technology support will have to take a more proactive role in serving the reminiscence functions and needs of older adults.  

%% file: sections/design.tex
\section{Concept of a Social Reminiscence Service}

We propose the design of an intelligent reminiscence chatbot that can play the role of \textit{companion} in reminiscence sessions with older adults. The chatbot acts as the guide of reminiscence session in a way that is engaging and fun, facilitating the tasks of collecting and organising memories and stories, while finding opportunities to connect people and stimulate social interactions. We exemplify these capabilities in Figure \ref{fig:bot-mockup}. 

The above requires the bot to adopt qualities that go beyond traditional 
scripted dialog systems \cite{ring2013addressing,demiris2016evaluation}, to qualities traditionally attributed to humans: i) understand \textit{triggers}, as to formulate meaningful questions in the same way someone looking at a picture would use that information to formulate related questions, ii) learn about the user, and use this \textit{knowledge} to elaborate more natural conversations and improve memory elicitation, iii) \textit{sense} where the conversation is going and guide the sessions accordingly, balancing the objective of collecting specific information (e.g., details on what a picture represents) with the opportunity to follow the person's flow of thinking and be directed by it, even if it takes us away from the picture we want to collect information on.  

Note that while use chatbot and text-based interactions in our example, these are only to illustrate the capabilities of the system, and to drive the concept development. Designing interactive technologies for older adults is a challenging task that requires specific attention to their abilities and limitations \cite{nurgalieva2017designing,hawthorn2000possible}. Thus, as part of part of our ongoing efforts we are investigating designs of conversational interfaces that can support individuals of potentially different interaction abilities. 









\subsection{Conceptual model}
The chatbot operates over a model comprising three families of concepts: the \textit{life model}, the \textit{reminiscence model}, and the \textit{conversation model}. 

The life model captures the information about the life of the person and the properties that describe them. 
This includes 
i) life events, which are well defined moments in the life of a person, such as birthdays, wedding, graduation, among others; 
ii) habits and skills, describing the person capabilities and routines, such as knitting or reading; 
iii) values, describing the person beliefs and ideals, such as religion, political views and social causes; 
iv) preferences, capturing passions and hobbies, such as music, sports, gardening; and v) relationships, referring to familial and non familial connections of the person. 
All this information is defined from a life course perspective, as they can refer to specific points in life. This temporal dimension cannot be measured in terms of specific dates, but in more flexible terms such periods of life.


\begin{figure*}
\centering
\includegraphics[width=.9\textwidth]{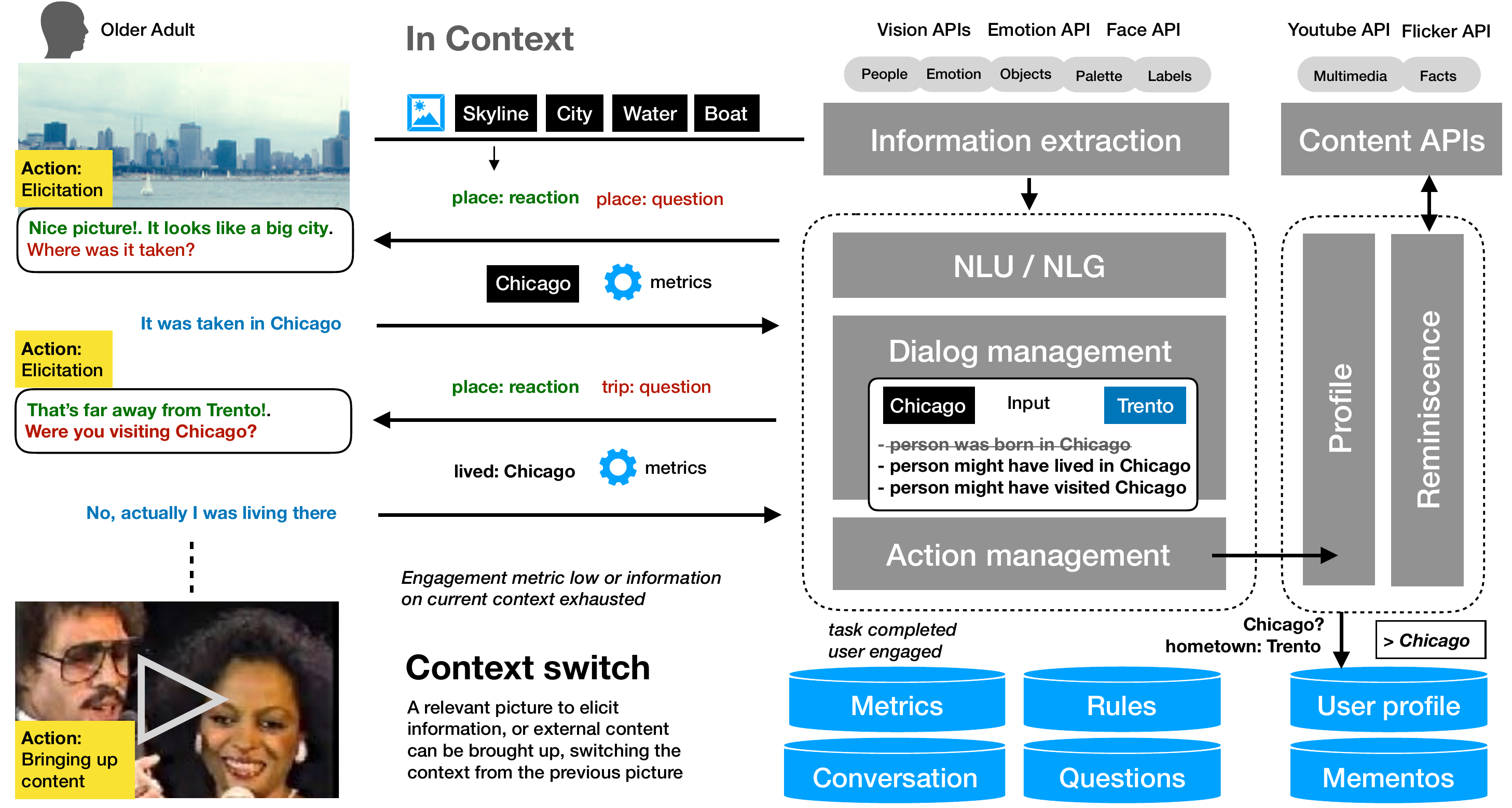}
\caption{Example of bot execution}
\label{fig:bot-exe}
\end{figure*}

Most of the concepts that are specific to the reminiscence practice were introduced in Figure \ref{fig:rem-process}. The reminiscence model defines: i) mementos,  digital artefacts (private or public) that can be used in the reminiscence session to trigger memories, such as pictures, videos or songs; ii) tags, annotations that describe the mementos, such as people, places, dates related to a picture; iii) life stories, anecdotes and memories that result from the reminiscence and storytelling session, possibly in response to a memento. Complementing this model there is the concept of the participant of the reminiscence session, which is captured by the life model, and the companion, which we define as part of the conversation model. 

The interactions between the user and the agent define \textit{conversations}. We organise conversations in \textit{sessions}, which give scope to the interactions during a reminiscence session. An interaction between the user and agent is usually referred to as \textit{turn}, and characterised by an input and a response. The set of input and responses define a \textit{context}, which provide implicit knowledge to the current and future turns. In the reminiscence bot, the memento under discussion, the questions recently asked, and the information provided by the user are key components of the immediate context. At a more global scale, the knowledge about the user (see life model) and the history of conversations define the \textit{macro context}. 
The \textit{actions} define the capabilities of our reminiscence bot, such as eliciting information, showing understanding (reactions), bringing up interesting multimedia content, and connecting people. Serving these actions we have a database of \textit{questions} related to specific user knowledge entities (to elicit information), a database of mementos (public and private, to bring up content), and a continuously updated database of user similarity and user clusters (to find potential connections). Orchestrating the decisions of what action to take and how, there is a set of \textit{rules}, and \textit{metrics} that facilitate the self-assessment of the bot.




\subsection{The chatbot at work}
The conversation model together with the reminiscence and life model work with the cognitive services as shown Figure \ref{fig:bot-exe}.

The picture used as trigger is first passed through a \textit{information extraction} service such as Google Cloud Vision that automatically detects its salient features (e.g., color palette, people, artefacts, landscape, emotions). In the example, the service detected a city "skyline" as the most important feature. Being the first interaction, and the context limited to the fact that a picture of a city was provided, the rule system decides to elicit basic information about the picture, embedding this information to elicit the related knowledge (\textit{place}: Nice picture! It looks like a big city. Where was it taken?). The user reply is then processed by an NLP service such as DialogFlow or Google Cloud Natural Language to extract the related entities. Metrics are also run to compute engagement and effectiveness of the action and question. The relevant tag (place: Chicago) is then associated to the picture.

In deciding the next action the system consider again the context, which is now comprised by a picture of Chicago, and information about the last exchange. This context combined with knowledge about the user (user: born in Trento) and metrics assessing a positive engagement, makes the following information elicitation rule to kick in ("User was born in Trento, and the picture is from Chicago: the user might have lived or visited Chicago"), requiring the connection of the user with the place to be inquired. This rule combined with the entity in context is paraphrased in a question as ("That's far away from Trento!. Were you visiting Chicago?). As before, the subsequent user responses are used to build the new knowledge that (The user had lived in Chicago in the 80s). Following the conversation, the interest of the user or the gain in eliciting information on the picture might decrease, motivating the bot to bring up new content related to the context to elicit new information, or suggest relevant connections.


It is important to note that the rules and context switching are not predefined, but can be trained using crowdsourcing and machine learning techniques, by crowds of experts and non experts depending on needs of the specific target population.

Mockups of the chatbot can be seen at \url{http://storygram.net}.

%% file: sections/architecture.tex
\subsection{Metrics and measures of success}



We briefly describe metrics that might shape the chatbot training and ability to self-assess:




\begin{easylist}
\ListProperties(Hide=100, Hang=true,Style*=\textbullet~)

& \textbf{Engagement} describes the level of involvement and interest of a person in the conversation. Capturing engagement is important as it has shown to be associated with longer conversations and higher sentiment score \cite{cervoneroving}.
Different engagement metrics have been proposed for conversational agents, such as: 
i) \textit{conversation length}, in minutes and/or amount of dialog turns \cite{fitzpatrick2017delivering,cervoneroving}; 
ii) \textit{total number of interactions} with the bot over a session \cite{fitzpatrick2017delivering};
iii) subjective \textit{user engagement ratings} \cite{cervoneroving}; and
iv) cumulative \textit{average sentiment score}, to see the user sentiment towards a topic \cite{cervoneroving}.


& \textbf{Task completion} captures the effectiveness achieved in the execution of a specific action.
The metrics that have been recently proposed for the task-based evaluation are: 
i) \textit{task completion rate} and ii) \textit{completion time}, which can be measured as a number of dialog turns the user talks to bot to complete a task \cite{huang2015guardian}.

& \textbf{Conversation quality} describes how much consistent the conversation is and the ability of the bot to correctly remember things the user said on the previous turns of conversation. For evaluation of conversation quality several metrics has been proposed: i) conversational consistency, ii ) memory of past events, and iii) response speed \cite{lasecki2013chorus}.

& \textbf{Human-like communication}
is proposed as a measure of the conversation quality and naturalness, where conversation logs are analysed to see if user had conversed with the bot as if with a human, e.g. using greetings or showing "proactivity" \cite{kopp2005conversational}.

\end{easylist}

On the other hand, measuring the actual benefits of the such a tool on the mental and social well-being of individuals would require the design of RCTs and the use of validated instruments.






%% file: sections/conclusion.tex
\section{Conclusion and Future work}


It is clear that conversational agents have the potential to overcome the limitations of traditional reminiscence technology, but further research is needed in order to go beyond scripted or human-operated systems. We did a step forward in this direction and proposed the use of well defined models (life, reminiscence and conversation models) in combination with cognitive services to provide agents with some human qualities.

The use of cognitive services is fundamental in providing bots with the required level of understanding to drive the reminiscence sessions. Recent developments in automatic caption and question generation \cite{mostafazadeh2016generating,tran2016rich} show promising results, and inspire our future work in building more intelligent reminiscence applications. 

As part of our ongoing and future work, we are investigating strategies for training the reminiscence bot with the help of the crowd. 